\documentclass{Interspeech2024}

\interspeechcameraready 

\title{The PESQetarian: \\On the Relevance of Goodhart's Law for Speech Enhancement}

\name{Danilo}{de Oliveira}
\name{Simon}{Welker}
\name{Julius}{Richter}
\name{Timo}{Gerkmann}

\address{Universit\"at Hamburg, Signal Processing}
\email{\{danilo.oliveira, simon.welker, julius.richter, timo.gerkmann\}@uni-hamburg.de}


\usepackage[dvipsnames]{xcolor}
\newcommand{\darkred}[1]{\textcolor{Maroon}{#1}}

\usepackage{flushend}
\usepackage{microtype}
\usepackage{adjustbox}
\usepackage{csquotes}
\usepackage{acronym}
\usepackage{cleveref}
\usepackage[subtle]{savetrees}
\acrodef{PESQ}{Perceptual Evaluation of Speech Quality}
\acrodef{POLQA}{Perceptual Objective Listening Quality Analysis}
\acrodef{ESTOI}{extended short-time objective intelligibility}
\acrodef{DNN}{deep neural network}
\acrodef{STFT}{short-time Fourier transform}
\acrodef{VB-DMD}{VoiceBank-DEMAND}
\acrodef{SNR}{signal-to-noise ratio}
\acrodef{WER}{word error rate}
\acrodef{SI-SDR}{scale-invariant signal-to-distortion ratio}
\acrodef{MOS}{mean opinion score}
\acrodef{MSE}{mean squared error}
\acrodef{SNR}{signal-to-noise ratio}

\begin{document}

\maketitle

\begin{abstract}
    To obtain improved speech enhancement models, researchers often focus on increasing performance according to specific instrumental metrics. However, when the same metric is used in a loss function to optimize models, it may be detrimental to aspects that the given metric does not see. The goal of this paper is to illustrate the risk of overfitting a speech enhancement model to the metric used for evaluation.
    For this, we introduce enhancement models that exploit the widely used PESQ measure. Our ``PESQetarian'' model achieves 3.82 PESQ on VB-DMD while scoring very poorly in a listening experiment. While the obtained PESQ value of 3.82 would imply ``state-of-the-art'' PESQ-performance on the VB-DMD benchmark, our examples show that when optimizing w.r.t. a metric, an isolated evaluation on the same metric may be misleading. Instead, other metrics should be included in the evaluation and the resulting performance predictions should be confirmed by listening.
\end{abstract}

\section{Introduction}

``When a measure becomes a target, it ceases to be a good measure'' \cite{strathern1997improving} --- originally stated to explain monetary policy \cite{goodhart1984problems}, Goodhart's law pertains to a much broader phenomenon which can be observed in various domains, from finance to education to natural sciences. In this paper we address the question whether this phenomenon also arises in speech enhancement.

Measuring success of a speech enhancement model is not straightforward, and developing instrumental metrics to measure success is a challenge by itself. There is a wide range of available metrics covering a variety of aspects, each with its strengths and weaknesses. Evaluation by means of listening experiments with human subjects is arguably the most reliable method for assessment of speech quality \cite{loizou2013speech}. But it comes with important inconveniences, such as its time-consuming nature and reproducibility issues, e.g. due to a limited number of participants or variations in the experimental conditions. To make the evaluation process faster and cheaper, instrumental objective metrics are continuously being developed. Focusing on speech quality, energy-based metrics \cite{hansen1998effective, vincent2006performance, roux2019sdr} are in general easy to compute, but do not model the processing steps of the human auditory system, therefore having a limited ability to predict subjective quality \cite{loizou2013speech}. As a consequence, much research effort has been directed to perceptually motivated measures, among which one of the most widely known and used is the \ac{PESQ} \cite{rix2001pesq} score. Sometimes this metric is even used to rank the performance of algorithms on a given dataset in order to define the current ``state-of-the-art'' in speech enhancement \cite{fu2021metricganp}.

Related to the question of how speech enhancement algorithms are evaluated is the question of how these systems are optimized. While the first approaches used energy-based optimization (e.g. the Wiener filter), already in 1985 one of the first attempts has been made to optimize perceptually more meaningful objectives in the log-amplitude domain \cite{ephraimMalah1985}. For a long time, researchers have attempted to use stronger perceptual metrics for optimization. For PESQ specifically, this is rather difficult due to the complex non-linear and non-differential operations involved in computing the metric. However, with modern machine learning techniques such an optimization became possible, either through reinforcement learning techniques \cite{koizumi2018dnn}, leveraging a \ac{DNN} to approximate \ac{PESQ} \cite{fu2019metricgan, fu2020learning, xu2022deep}, or by slightly modifying PESQ to make it differentiable \cite{donas2018deep, kim2019end}.

Nevertheless, in light of Goodhart's law, isn't it problematic to optimize an algorithm on the same objective function used for evaluating the algorithm? This is the research question we aim to address in this paper.

While several works optimize their models on the \ac{SI-SDR} \cite{luo2019conv}, we focus on optimizing \ac{PESQ} and demonstrate how it can be wrongly exploited by a \ac{DNN}. Since PESQ was not developed for the evaluation of speech enhancement models, especially non-linear ones, using it as an optimization goal can have unexpected and undesired effects, such as artificial distortions that the measure was not designed to penalize. A mismatch between PESQ and other metrics has been found by \cite{bie2022unsupervised, richter2023speech} to happen when PESQ is used as the optimization criterion, e.g. as proposed in \cite{fu2021metricganp}. In this work, we perform a detailed analysis of this phenomenon. We compare the same network architecture trained with different loss functions, including one with an adversarial term that aims at maximizing distortions, and show that good \ac{PESQ} scores can be obtained to the detriment of listening experience. We leverage the findings from our trained \acp{DNN} to further investigate the mechanics through which the measure can be tricked. Concurrent work by \cite{close2024hallucination} shows that similar phenomena also occur in the case of non-intrusive metrics such as DNSMOS \cite{reddy2022dnsmosp835}. The results of our \textit{PESQetarian} models highlight the relevance of Goodhart's law also to speech enhancement. Our outcomes emphasize the risks of blindly trusting and optimizing for instrumental metrics, and highlight the importance of a complete evaluation process to assess the performance of speech enhancement models.

\section{Speech enhancement and PESQ}

The task of speech enhancement consists in processing speech signals degraded by additive noise to improve perceptual aspects, such as quality and intelligibility \cite{loizou2013speech}. In deep learning-based speech enhancement, some or even all of the processing steps are carried out by a \ac{DNN}, leveraging the powerful capabilities these have of learning complex non-linear functions from data. Given a mixture containing additive noise and a clean speech utterance $\mathbf{s}\in\mathbb{R}^N$, where $N$ is the number of samples, the model is tasked with outputting a clean speech estimate $\mathbf{\hat{s}}\in\mathbb{R}^N$.

PESQ is a full-reference algorithm for the automated assessment of speech quality in telecommunication systems, predicting the \ac{MOS} score that would be assigned to the degraded signal. In 2001 it was officially standardized as ITU-T Recommendation P.862 \cite{itu-t-p-862}. It is important to note that this recommendation has already been withdrawn, being superseded by \ac{POLQA} \cite{beerends2013perceptual} (ITU-T Recommendation P.863 \cite{itu-t-p-863}). However, PESQ remains the de facto standard metric for instrumental measurement of speech quality in speech enhancement research.

The PESQ algorithm pipeline starts with level and time alignment of reference and degraded signals. The level-aligned signals are then processed using a perceptually-motivated auditory transform. Differences (disturbances) are aggregated in frequency and time, being finally mapped to the MOS scale. Although PESQ is originally not differentiable, there exist implementations that perform slight modifications to make it differentiable and usable as a loss function \cite{donas2018deep, kim2019end}. PESQ has also been included in the training process of deep speech enhancement models via reinforcement learning techniques \cite{koizumi2018dnn} and more recently through adversarial training \cite{fu2019metricgan, fu2021metricganp, cao2022cmgan}.

\section{Experimental setup}

\subsection{Loss function}

As a baseline, we use a scaled \ac{MSE} loss in the time domain, which works purely at the signal level and does not include perceptual components:

\begin{equation}
    \mathcal{L}_\text{MSE}(\mathbf{s}, \mathbf{\hat{s}}) = \frac{1}{2}\sum_{i=1}^N(s_i - \hat{s}_i)^2
\end{equation}

Since PESQ is not differentiable, we use an alternative differentiable implementation for PyTorch\footnote{\url{https://github.com/audiolabs/torch-pesq}}, called torch-pesq. This implementation is based on \cite{donas2018deep} and \cite{kim2019end}. It dispenses with time alignment, since the enhancement model already outputs aligned utterances, and performs level alignment with infinite impulse response (IIR) filtering instead of frequency weighting. The package provides a class for the use of PESQ as a loss function $\mathcal{L}_\text{torchPESQ}$, so that minimizing $\mathcal{L}_\text{torchPESQ}$ approximately means maximizing PESQ.

In order to investigate how loopholes in the PESQ computation could be exploited by an algorithm optimizing for it, we additionally experimented with adding the \ac{SI-SDR} \cite{roux2019sdr} metric as a loss function term. We aim at minimizing the \ac{SI-SDR}, therefore maximizing the distortion. The goal here is to encourage the model to find ways to get good PESQ scores at unreasonably distorted conditions. We add hyperparameters $\alpha$ and $\beta$ to balance the two loss terms. $\beta$ scales the \ac{SI-SDR} term to the same magnitude as the torchPESQ term, while $\alpha$ controls which term will have a larger weight. The resulting loss is
\begin{equation}
    \mathcal{L}(\mathbf{s}, \mathbf{\hat{s}}) = \alpha\text{ }\mathcal{L}_\text{torchPESQ}(\mathbf{s}, \mathbf{\hat{s}}) + (1-\alpha)\text{ }\beta\text{ }\mathcal{L}_\text{SDR}(\mathbf{s}, \mathbf{\hat{s}}).
\end{equation}

\subsection{Model}
As a \ac{DNN}, we employ the NCSN++ architecture \cite{song2021scorebased}. It was successfully used as a backbone for generative diffusion models \cite{richter2023speech}, and was shown to have good denoising performance also as a discriminative model \cite{lemercier2023analysing}. We use a publicly available codebase\footnote{\url{https://github.com/sp-uhh/sgmse}}, adapting it to the discriminative training setting by performing complex spectral mapping. It is a multi-resolution U-Net with residual blocks containing mainly filters performing up- or downsampling and 2D convolutional layers. There are channel-wise attention blocks in the bottleneck and at the $16\times16$ resolution. Additionally, a progressive branch performs down-/upsampling using shared weights, with a contracting path providing information to the U-Net's encoder layers and an expansive path incorporating information from the decoder to produce the output. 

The inputs to the model are complex spectrograms, with the real and imaginary parts stacked along the channel dimension. For this, a \ac{STFT} operation is performed, using a 510-point FFT with a hop length of 128 and a Hann window of size 510, resulting in 256 frequency bins. Inspired by \cite{lemercier2023analysing}, we also use a reduced version of the original architecture, reducing the original number of residual blocks per stage from two to one, and slightly reducing the number of blocks: 5 residual block layers in the encoder, with output channels $[128, 128, 128, 256, 256]$. The resulting model contains approximately $30\,$M parameters. We train with the Adam optimizer \cite{kingma2015adam}, using batch size $16$ and learning rate $ 10^{-4}$. We train for a maximum of 200 epochs and select the models with the best PESQ scores on the validation set. In the PESQ-SDR experiment, we set the loss weights to $\alpha=0.9$ and $\beta=0.5$.

\subsection{Dataset and evaluation}\label{section:evaluation}

The data used to train the models comes from the publicly available \ac{VB-DMD} dataset \cite{valentini2016investigating}. We use the audio downsampled to 16$\,$kHz. The training set contains speech mixed with either real noise recordings or artificially-generated noise at \ac{SNR} levels of 0, 5, 10 and 15dB. The test set contains mixtures with unseen real noise types at 2.5, 7.5, 12.5 and 17.5$\,$dB \ac{SNR}. Two speakers from the training set are left out as a validation set.

For instrumental evaluation of speech quality, we make use of PESQ, POLQA and SI-SDR \cite{roux2019sdr}, additionally employing DNSMOS OVRL \cite{reddy2022dnsmosp835} as a reference-free metric. We evaluate the \ac{ESTOI} measure \cite{taal2011algorithm} for intelligibility, and additionally \ac{WER} for assessment of speech recognition performance. \ac{WER} is computed on transcriptions of clean and corresponding model enhanced estimates. The transcriptions are automatically generated by QuartzNet \cite{kriman2020quartznet}. 

We also conduct a listening experiment with 10 audio experts, asked to rate 10 random samples from the test set that are longer than 4 seconds. Each sample contained 5 randomly ordered and unidentified candidates: enhanced estimates by the 3 candidate models, plus the reference and the noisy mixture as an anchor. The ratings are given in a scale from 0 to 100. In order for the PESQ-SDR model to be audible, we processed all files, setting the two lowest frequency bins to zero and cropping out the first 0.5 second of audio. This successfully removed any loud clicks from the model output samples (see Section~\ref{section:results}).

\section{Results and discussion}

\begin{table*}[t]
\adjustbox{max width=\textwidth}{
    \begin{tabular}{@{}ll|cccccc@{}}
        \toprule
         Name &Objective& PESQ & POLQA (Support) & SI-SDR [dB] & ESTOI & DNSMOS OVRL & WER [\%] \\
        \midrule
        Noisy & --- & $1.97 \pm 0.75$ & $3.11 \pm 0.79\text{ }(57\%)$ & $8.4 \pm 5.6$ & $0.79 \pm 0.15$ & $2.69 \pm 0.53$ & $8.3 \pm 16.1$ \\
        MSE &MSE$\downarrow$ & $2.78 \pm 0.74$ & $\mathbf{3.87 \pm 0.56}\text{ }(56\%)$ & $\mathbf{18.8 \pm 3.3}$ & $\mathbf{0.87 \pm 0.10}$ & $\mathbf{3.13 \pm 0.23}$ & $\mathbf{7.6 \pm 15.8}$ \\
        \midrule
        PESQetarian &PESQ$\uparrow$ & $\mathbf{3.82 \pm 0.57}$ & $\darkred{1.46 \pm 0.48}\text{ }(56\%)$ & $\darkred{-19.8 \pm 3.3}$ & $0.84 \pm 0.09$ & $\darkred{2.06 \pm 0.46}$ & $\darkred{8.5 \pm 16.7}$ \\
        PESQ-SDR&PESQ$\uparrow$, SDR$\downarrow$ & $3.26 \pm 0.40$ & $\darkred{1.44 \pm 0.31}\text{ }(41\%)$ & $\darkred{-70.6 \pm 10.2}$ & $\darkred{0.72 \pm 0.17}$ & $\darkred{1.70 \pm 0.39}$ & $\darkred{60.4 \pm 37.6}$ \\
        \bottomrule
    \end{tabular}}
    \caption{Enhancement results on VB-DMD. Arrows indicate whether the model is trained to maximize or minimize the loss function terms. Best values are represented in bold, values worse than for Noisy in red. \textit{Support} is the percentage of files that met the POLQA criteria.}
    \label{tab:results}
\end{table*}

\begin{figure}
    \centering
    \includegraphics[width=0.8\linewidth]{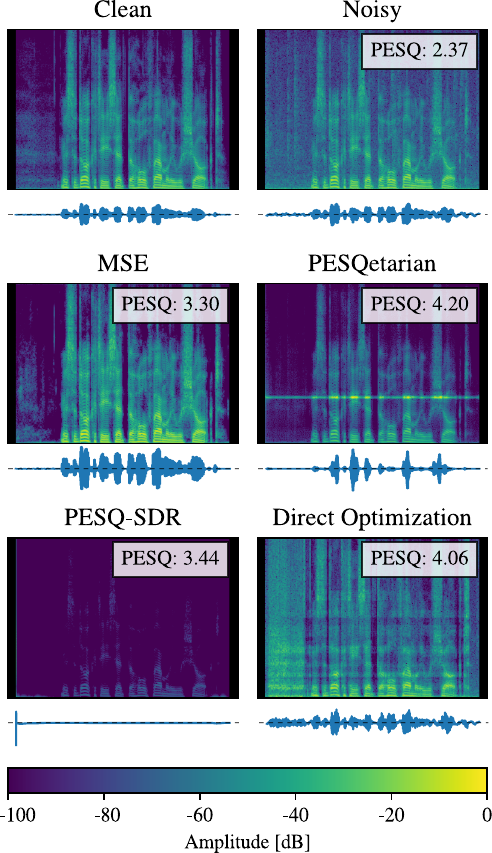}
    \caption{Spectrograms of estimates produced by the models for a given utterance, accompanied by the corresponding PESQ values. The spectrograms were padded to allow for visualization of the click produced by the PESQ-SDR model.}
    \label{fig:example}
\end{figure}

\subsection{The PESQetarian}\label{section:results}

Table~\ref{tab:results} shows the instrumental evaluation of our models trained on the different loss functions. Under ``Support'' we provide the percentage of the test set over which \ac{POLQA} is computed successfully due to requirements of the metric, e.g., minimum of 3 seconds of active speech. It is interesting to note that while existing PESQ implementations also allow processing shorter signals, according to the P.862.3 standard~\cite{itu-t-p-862-3} there should actually be a minimum of 3.2s active speech in the reference. As expected, the PESQetarian model obtains the best PESQ scores, followed by the PESQ-SDR model. The PESQetarian obtains 3.82 in PESQ, which would imply ``state-of-the-art''. However, almost all other metrics are worse than for the noisy mixtures, demonstrating that the optimization process successfully exploits PESQ in detriment of other metrics. The MSE model obtains the best results overall for all other metrics. Since optimizing for \ac{MSE} is related to optimizing for \ac{SI-SDR} \cite{heitkaemper2020demystifying}, it is no surprise that the \ac{MSE} model has a particularly good score in that measure.

\begin{figure}
    \centering
    \includegraphics[width=0.9\linewidth]{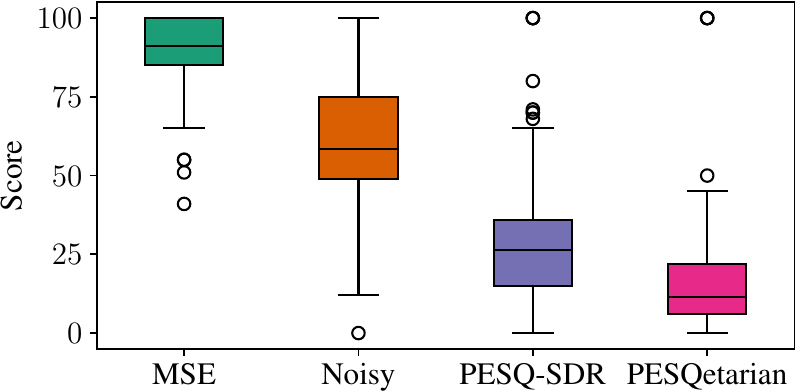}
    \caption{Results of the listening experiment. Due to the \enquote{clicks} introduced by the PESQ-SDR model (see Section~\ref{section:clicktrick}), the audio was processed to make the utterances audible, as described in Section~\ref{section:evaluation}.}
    \label{fig:mushra}
\end{figure}

A quick listening check of the model outputs reveals that the PESQetarian model produces high pitched artifacts, while the PESQ-SDR model introduces a very loud click in the beginning of the audio, making the rest of the utterance inaudible due to the excessive dynamic range (see Figure~\ref{fig:example}). It is worth to note that the estimates produced by the PESQ-SDR model only have negative values and thus do not have mean zero, as seen in Figure~\ref{fig:example}. 
The outcomes of the listening experiment are presented in Figure~\ref{fig:mushra}. The results reinforce the superiority of the MSE model, with the others being perceived as worse than the noisy mixture. The PESQetarian provides the worst listening experience, likely due to the high-pitched artifacts it introduces. It is interesting to note that the processed PESQ-SDR outputs also exhibit quantization noise, as the \enquote{click} drastically increases the dynamic range of the audio signal. Please note that which loopholes of PESQ are exploited --- and to what extent --- depends on the loss weights $\alpha$ and $\beta$ as well as the random \ac{DNN} weight initialization, and these effects are not guaranteed to happen.

\subsection{A click trick}\label{section:clicktrick}
In Section~\ref{section:results}, we found that the trained PESQ-SDR model inserts a short click at the beginning of the utterances. To see whether this is only due to SI-SDR minimization or due to PESQ maximization, we investigated what PESQ scores can be achieved simply by replacing the first sample in the noisy utterances by a fixed large value $c$ (the \emph{click value}), i.e., $\hat{s}_0 \gets c\,$, 
where all other $\hat{s}_i, i > 0$ are copied from the noisy audio.

We determined which value of $c$ can lead to optimal PESQ improvements on the VB-DMD set, and found a median optimal click value of $c = 666$. This one-parameter \enquote{model}, when applied to \enquote{enhance} each noisy file, achieves a PESQ of ${3.46\,\pm\,0.71}$ on the VB-DMD test set, which would be fourth best on the leaderboard\footnote{\url{https://paperswithcode.com/sota/speech-enhancement-on-demand}} for this task as of June 2024.

The ITU-T Recommendation P.862.3 \cite[Sec. 7.8]{itu-t-p-862-3} notes that \enquote{a minimum leading and trailing silence of 0.5 s is recommended} when applying PESQ. Our discovered trick of inserting this click completely ignores this recommendation, instead exploiting the resultant detrimental effects on the PESQ system. Of course, we do not mean to present this as an actual enhancement method, but rather to draw attention to the fact that such intricacies and loopholes of specific metrics can be found and exploited effectively by deep learning: our PESQ-SDR model independently discovered this trick not from a manual white-box analysis of PESQ, but purely through the training process.

Nonetheless, we conducted such a white-box analysis ourselves using the source code of the torch-pesq Python package. We found the main reason that a simple click can lead to high PESQ scores lies in the level alignment, where, among other quantities, the power (sum of squares) of the reference and degraded signal is determined. This quantity is highly sensitive to outliers such as a large-valued click, leading to a erroneous level estimation for the degraded signal and resulting in severe underestimation of the disturbance. Additionally, the click itself -- when placed at the signal index 0 -- does not enter the disturbance estimate in the subsequent processing steps of PESQ that are based on the \ac{STFT}, since it is multiplied with 0 by the employed Hann window function.

We found empirically that by replacing the sum of squares in the level estimation by the 85th percentile of the squared signal samples, scaled up by the signal length, one can make torchPESQ essentially immune to the \enquote{click trick} without sacrificing alignment with real PESQ scores (Pearson correlation of $r = 0.973$ compared to PESQ on the VB-DMD test set). This finding suggests that metrics can be tested for weak points and potentially made more robust. Nevertheless, other loopholes can still be found and exploited by a \ac{DNN}, which is the case of our PESQetarian model.

\begin{figure}
    \centering
    \includegraphics[width=\linewidth]{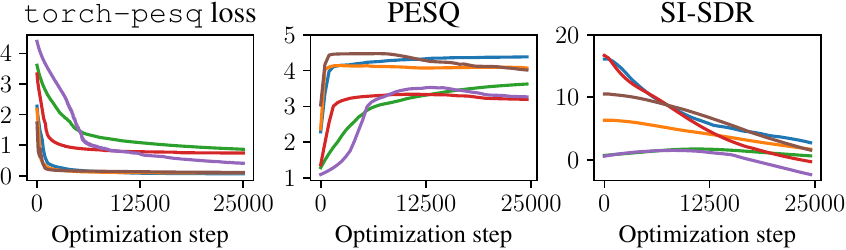}
    \caption{torchPESQ loss, PESQ, and SI-SDR for each iteration of the single-utterance oracle PESQ optimization procedure. Each line represents one utterance. While large gains in PESQ are achieved, SI-SDR is consistently significantly worsened.}
    \label{fig:oracle-pesq-optimization}
\end{figure}

\begin{table}[h]
    \centering
    \begin{tabular}{@{}l|cc@{}}
        \toprule
        Metric & Noisy & Optimized \\
        \midrule
        torchPESQ & $1.89 \pm 0.75$ & $\mathbf{4.40 \pm 0.25}$ \\
        PESQ & $1.91 \pm 0.77$ & $\mathbf{3.76 \pm 0.47}$ \\
        POLQA & $\mathbf{3.46 \pm 0.21}$ & $2.73 \pm 0.69$ \\
        SI-SDR & $\mathbf{8.49 \pm 7.17}$ & $0.54 \pm 1.78$ \\
        ESTOI & $\mathbf{0.80 \pm 0.19}$ & $0.73 \pm 0.07$ \\
        DNSMOS OVRL & $\mathbf{2.63 \pm 0.51}$ & $2.08 \pm 0.32$ \\
        \bottomrule
    \end{tabular}
    \caption{Results of our oracle PESQ optimization procedure (see Section \ref{section:oracle-optim}) on six VB-DMD test files.}
    \label{tab:oracle-pesq-optimization}
\end{table}

\subsection{Single-utterance oracle PESQ optimization}\label{section:oracle-optim}
To show that metrics such as PESQ can be exploited not only by \acp{DNN} but also by simple optimization, we conduct an additional experiment treating each sample $\hat{s}_i$ contained in the audio estimate $\mathbf{\hat s}$ as a free parameter to optimize in order to achieve a maximum PESQ value compared to the clean audio $\mathbf s$. For each specific utterance $\mathbf{s}$ we consider the optimization task:
\begin{equation}\label{equation:max-oraclepesq}
    \min_{\mathbf{\hat s}}\ \mathcal{L}_{\text{torchPESQ}}(\mathbf s, \mathbf{\hat s})
\end{equation}

Note that in contrast to the DNN training where we optimize the parameters of the DNN, here we optimize $\mathbf{\hat s}$ itself. We initialize $\mathbf{\hat s}$ with the noisy audio samples, and run the Adam optimizer \cite{kingma2015adam} for 25,000 iterations with a learning rate of $10^{-5}$. This treats $\text{PESQ}(\mathbf s, \cdot)$ as a differentiable black-box that has all information about $\mathbf s$ available (in the form of $\mathbf s$ itself), but only provides it to the optimizer through the view of the metric. Thus any errors to which PESQ is blind will be undetected by the optimization procedure, and any artifacts that somehow improve PESQ scores will be amplified. To avoid that this method employs the \enquote{click trick} as discussed in Section~\ref{section:clicktrick}, we keep the first and last few hundred samples unchanged during the optimization. 

We perform the optimization method described above on six random utterances from VB-DMD. We show the results compared to the noisy files in Table \ref{tab:oracle-pesq-optimization}. In Figure \ref{fig:oracle-pesq-optimization}, we further show PESQ and SI-SDR for all iterations of the optimization process. We find that the method successfully reduces the torchPESQ loss and significantly improves PESQ scores for all utterances, but at the same time reliably worsens SI-SDR and all other evaluated metrics. This shows that PESQ and SI-SDR are potentially at odds with each other, suggesting that when employing combined loss functions optimizing for, e.g., both PESQ and SI-SDR, the involved relative weighting must be carefully tuned. While the PESQ values are improved significantly for most of the optimization procedure, for some utterances they are eventually also worsened again while the loss is still being reduced. This points to \enquote{overfitting} on the torchPESQ loss, where even minor implementation details or differences of torch-pesq seem to be exploited to the detriment of PESQ scores.

An informal listen reveals that the files do not sound as good as their PESQ values ($3.76\,\pm\,0.47$) would suggest, containing distortions as well as strong noise components with increased amplitude compared to the noisy files. We also provide the \enquote{enhanced} files from this method on our demo website\footnote{\url{https://uhh.de/inf-sp-pesqetarian}}.

\section{Conclusions}

This paper presented a study on optimizing speech enhancement models for the popular PESQ metric. It is important to point out that the models developed in this paper are only for research purposes and are not to be used in practical applications.
We showed that such optimization with respect to PESQ can lead to absurd results where processed speech has a very good PESQ score but receives a very low score in listening experiments. To further analyze the issue, we trained a model to jointly maximize PESQ and minimize SI-SDR, revealing one of the tricks that can lead to a spuriously high PESQ score --- a loud click in the beginning of the audio. Our results illustrate the risks of overfitting to a given instrumental metric and highlight the necessity of evaluating speech enhancement models on a variety of metrics, as well as the paramount importance of formal listening experiments. Furthermore, these results demonstrate how DNNs can be used to analyze and validate instrumental metrics. 

\section{Acknowledgements}
This work was funded by DASHH (Data Science in Hamburg - HELMHOLTZ Graduate School for the Structure of Matter) - Grant- No. HIDSS-0002, and by the German Research Foundation (DFG) in the transregio project Crossmodal Learning (TRR 169). We would like to thank J. Berger and Rohde\&Schwarz SwissQual AG for their support with POLQA.

\bibliographystyle{IEEEtran}
\bibliography{mybib}

\end{document}